\documentclass[twocolumn,floats,prl,aps]{revtex4}

\usepackage[dvips]{graphicx}
\usepackage{amsfonts}
\usepackage{amsmath}
\usepackage{amssymb}
\usepackage{exscale}
\usepackage{eufrak}

\begin{document}

\draft    % Show PACS

%\wideabs{    %>>> Not to use in preprint mode.

\title{Optical Spectroscopy as a Probe of Gaps and Kinetic Electronic Energy  in p- and n-type cuprates}


\author{Nicole Bontemps}

\address{Physique du Solide-UPR5 CNRS, ESPCI 10 rue Vauquelin, 75005 Paris, France}

\begin{abstract}
The real part of the optical in-plane conductivity of p-- and n--type cuprates thin films at various doping levels  
was  
deduced from highly accurate reflectivity measurements. We present here a comprehensive set of 
 optical spectral weight data as a function of the 
temperature $T (> T_c$), for underdoped and  overdoped samples. The temperature dependence of the  spectral weight is not universal. Using various cut-off frequencies for  the  spectral  
weight, we show that n--type Pr$_{2-x}$Ce$_x$CuO$_4$  and 
 p--type Bi$_2$Sr$_2$CaCu$_2$O$_{8+\delta}$ exhibit both similarities 
and striking differences. 
 The Fermi surface is closed in overdoped metallic samples. In underdoped Pr$_{2-x}$Ce$_x$CuO$_4$ samples, it clearly 
 breaks into arcs, giving rise to a "pseudogap" 
 signature.  It is argued that such a 
 signature is  subtle in underdoped  Bi$_2$Sr$_2$CaCu$_2$O$_{8+\delta}$.  

\end{abstract}

\maketitle  %>>> To use with RevTeX4.  Not to use with standard RevTeX.

% main text
\section{Introduction}
%\label{1}

The properties of a doped Mott insulator and the role of strong correlations stand among the present great 
challenges in condensed matter physics, and high critical temperature superconductivity in cuprates is one of those.  
Unlike conventional metals, various energy scales are present in the normal state of these materials, 
 and vary throughout the phase diagram. 
 
 One energy scale is related to the pseudogap, 
evidenced by many experimental techniques \cite{Timusk}.
Angular Resolved Photoemission Spectroscopy (ARPES) of hole doped cuprates  
 yields a "low energy"  
pseudogap, opening below a (doping dependent) temperature $T^{\star}$. This pseudogap is traced 
by the shift with respect to the 
 Fermi level, of the leading edge of the energy dependent curve (EDC)  
 along  the ($0, \pi$) direction of the {\bf k} space.    
It is argued that a "high energy" pseudogap develops, also along the ($0, \pi$) direction, 
associated with a broad feature which tracks the low energy pseudogap \cite{ZXrev}. 
 As for infrared spectroscopy, which is the topic of this paper, hole doped cuprates reveal a pseudogap 
 feature only through the depletion of the low energy optical  scattering rate  \cite{Puchkov96}.
 
In electron-doped cuprates,  a "low energy" 
pseudogap  was observed by tunneling spectroscopy \cite{UM}. 
A "high energy" pseudogap is seen in ARPES also as a broad feature in the EDC  \cite{Matsui05}, however  
 along the ($ 0.65 \pi, 0.3 \pi$) direction. The infrared in-plane conductivity reveals a pseudogap at 
 a comparable energy scale.
 The temperature $T^{\star}$  where this "pseudogap"  opens depends on doping  
 \cite{Onose04,Zimmers05}. 
 This doping dependence  of T$^{\star}$ is best tracked, as shown further,  by measuring the 
spectral weight $W (\Omega, T)$, 
 {\it i.e} the integral of the real part $\sigma_{1}(\omega)$ of the optical conductivity up to 
 some cut-off frequency $\Omega$:

\begin{equation}
	W(\Omega,T)=\int_{O}^{\Omega} \sigma_{1}(\omega, T) d\omega 
\label{SW}
\end{equation}

Another energy scale is relevant to cuprates, namely the electronic kinetic energy $E_{K}$.
In a single band,  tight-binding model with nearest-neighbor interactions,  $E_{K}$ is related to the spectral 
weight, provided the cut-off frequency $\Omega$ is chosen so as to restrict to the free carrier contribution 
\cite{NBAnnals}.
The electronic kinetic energy scale 
reflects the strength of the correlations \cite{Millis05}. 
Its temperature dependence remains unsettled, from a theoretical point of view and  
from  the experiments where the temperature dependence  varies from sample to sample 
\cite{Benfatto06,Ortolani05,Carbone06,Hwang06}. 
Such discrepancies suggest that the $T$ dependence of $E_{K}$ (or of the spectral weight) is not universal. 

In this communication, we take advantage of our measurements in both n--type Pr$_{2-x}$Ce$_x$CuO$_4$ (PCCO) and 
 p--type Bi$_2$Sr$_2$CaCu$_2$O$_{8+\delta}$ (BSCCO) thin films. Relying on the metallic behavior of the overdoped 
samples, whether n-- or p--type,  we note that the 
 underdoped compounds deviate from this metallic behavior at some temperature T$^{\star}$. We argue that 
 such deviations, although quite different in PCCO and BSCCO, are associated with the onset of a "pseudogap". 
 One caveat though: we use the term "pseudogap" for the sake of simplicity, and we mean a density gap.
 However it may relate to different physics . This   
still  vividly debated issue  is beyond the scope of this communication.

\section{Experimental}
%\label{2}
Samples are thin films with thicknesses ranging from 2000 to 3000~\AA. 
Details about their preparation and characteristics and about the experimental procedure  can be found elsewhere 
\cite{Santander04,Zimmers05}.

%-------------
\begin{figure}
 % \begin{center}
      \includegraphics[scale = 0.45]{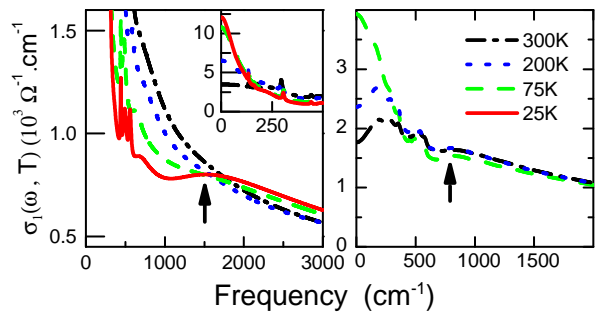}
 % \end{center}
  \caption{ Left panel: real part $\sigma_{1} (\omega,T)$ of the optical conductivity of the underdoped 
  PCCO sample     -- T$_c$=15~K. The inset shows the conductivity in the range 
  0-500~cm$^{-1}$ over the full scale along the Y scale, an expanded view 
   in the main left panel shows the changes of $\sigma_{1} (\omega)$ with temperature over the range 
   0-3000~cm$^{-1}$.The arrow shows a clear maximum in the PCCO conductivity at 25~K, just above T$_c$. 
  Right panel: real part $\sigma_{1} (\omega,T)$ of the optical conductivity of the underdoped 
  BSCCO sample -- T$_c$=70~K. $T$~=~300~K,~200~K,~75~K,~25~K (the latter lying below $T_c$ for BSCCO is 
  not shown).  The arrow shows a tenuous maximum in the BSCCO conductivity at 75~K, just above T$_c$.}
  
  \label{Fig1}
\end{figure}
%-------------

Figure 1 displays the optical conductivity for two  PCCO and BSCCO underdoped samples 
 in a temperature range where pseudogaps were claimed previously to be identified. 
A clear dip and a maximum at $\sim$~1700~cm$^{-1}$ develop in the PCCO sample as  
$T$ decreases. A dip and a maximum at $\sim$~800~cm$^{-1}$ are hardly visible at 75~K
in the case on underdoped BSCCO. 
If such maxima relate to a density gap feature, then upon the opening of such a gap, 
states are transferred from the Fermi  energy to a higher energy.  
 This reflects as a spectral  weight transfer toward high energy at the temperature where the gap opens. 
 We expect this transfer of spectral weight to be best observed when integrating up to 
 a cut-off frequency $\Omega$ (Eq.1) of the order of the gap energy. 
 
Figures 2 and 3 show the change of spectral weight for a set of cut-off frequencies spanning 
the range $\Omega = 500-20000$~cm$^{-1}$ in PCCO (fig.2) and BSCCO (fig.3).   
The  $\Omega$ cut-off frequencies are the same in a single family,
but not in-between the two families. The  values were  selected 
in order to display the characteristic energy scales where changes in the temperature dependence are best seen. 

We  translated the spectral weight in units of a carrier 
density per copper $\rm N_{eff}$ \cite{Hwang06}. An estimate of the actual effective carrier density per copper 
is obtained for $\Omega=8000$~cm$^{-1}$. We find $\rm N_{eff}(300K)$ =0.22 and 0.26 for the 
overdoped and underdoped PCCO respectively, 0.32 and 0.35  for the underdoped and overdoped BSCCO respectively 
\cite{Neff}.
The difference ${\rm N}-{\rm N_{eff}(300~K)}$ versus $T$ is displayed in fig.2 and 3. The scatter of the points 
at 20000~cm$^{-1}$ in the right panel of fig.2 gives an upper limit for our data accuracy  
($\pm~0.002$ in units of $\rm N_{eff}$) \cite{note2}.

%-------------
\begin{figure}
 % \begin{center}
      \includegraphics[scale = 0.43]{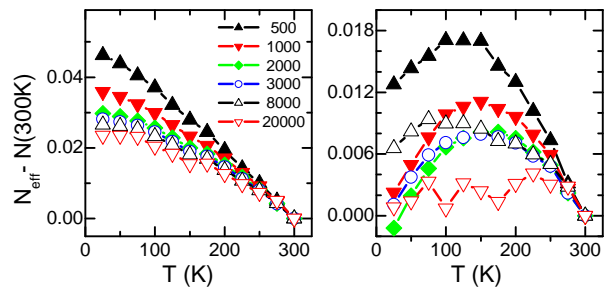}
 % \end{center}
  \caption{ Change of spectral weight  -- translated into the effective 
  carrier density  $\rm N_{eff}$ per Cu -- versus temperature, for the underdoped -- x=0.13, $T_C$=15~K -- 
  and   overdoped -- x=0.17, $T_C$=15~K --   PCCO films. 
    The cut-off frequencies $\Omega$ are indicated in the left panel. }
  \label{Fig2}
\end{figure}

%-------------
\begin{figure}
 % \begin{center}
      \includegraphics[scale = 0.43]{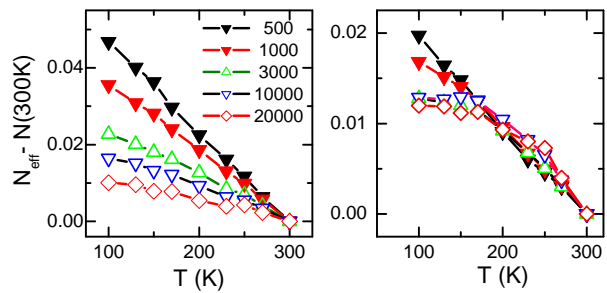}
 % \end{center}
  \caption{ Change of spectral weight (SW) -- translated into the effective 
  carrier density  $\rm N_{eff}$ per Cu -- versus temperature, for the underdoped -- $T_c$=70~K -- and
  overdoped -- $T_c$=63~K --    BSCCO films. The cut-off frequencies $\Omega$ are indicated in the left panel.} 
    
  \label{Fig3}
\end{figure}
%-------------

Both overdoped samples (left panels, fig.2 and 3) exhibit a similar behavior: ${\rm N}-{\rm N_{eff}(300~K)}$ 
increases steadily as $T$ 
decreases, approximately by the same amount.
This increase  results from a single mechanism, which does not depend upon 
 the details of the Fermi surface:  the quasiparticle (QP) scattering rate is overall decreasing as the 
 temperature decreases, thus putting more and more spectral weight in the Drude-like peak at low energy. 
${\rm N}-{\rm N_{eff}(300~K)}$ is the smallest at the largest cut-off; this is the expected trend since the 
f-sum rule must be eventually satisfied at large frequencies. Note that in the case of BSCCO, 
at 20000~cm$^{-1}$, ${\rm N}-{\rm N_{eff}(300~K)}$ lies indeed almost within the experimental  
error, while the f-sum rule is not fulfilled  yet at this energy in PCCO. 

The point we wish to emphasize is that the overdoped samples exhibit a metallic behavior, associated 
with a closed Fermi surface, and there is no hint as T decreases, of a modification of this Fermi surface.

In underdoped PCCO (fig.2, right panel), as $T$ is lowered from 300~K, the spectral weight increases as 
observed in the overdoped samples: this is a signature  of a metallic state. The material remains 
metallic down to 25K,
as certified by a  Drude-like peak (shown in the inset of fig.1) in $\sigma_{1} (\omega,T)$.
Below $\sim$~150~K, the trend is reversed: ${\rm N}-{\rm N_{eff}(300~K)}$ decreases.  
This can be understood without 
the need of a specific model. The optical conductivity is an average over the {\bf k} space. 
Since a metallic behavior is present down to 25~K, the observed decrease implies that 
the Fermi surface breaks into arcs, possibly due  to a  partial gap (pseudogap) opening along
 specific $\bf k_0$ values (identified by ARPES).  The QP states 
which  vanish at those $\bf k_0$ values no longer contribute  to the Drude peak, and are transferred 
at high energy, hence a  decrease of the low energy spectral weight below 150~K. 
The characteristic energy scale of this  pseudogap  lies in the range 2000-3000~cm$^{-1}$. 
It corresponds to  the maximum seen in the optical conductivity at $\sim$~1700~cm$^{-1}$ 
(fig.1, left panel). It is also is compatible with ARPES data \cite{Matsui05}.

The spectral weight in the underdoped BSCCO sample does not decrease whatever the cut-off 
frequency  although it was independently experimentally  established  that a pseudogap is present 
 below a temperature T$^{\star}$ in this material  (in this specific sample, 
 from the resistance measurement, $T^{\star}$~$\sim$~130~K \cite{NBAnnals}).
 Nevertheless,  figure 3 (right panel) shows that, at $\Omega \simeq 3000$~cm$^{-1}$, the spectral
 weight levels off. The temperature where this levelling off occurs is $\sim$~140~K. 
 ${\rm N}-{\rm N_{eff}(300~K)}$ remains constant 
 down to $T_c$, showing  
  no  decrease as compared to underdoped PCCO. Moreover, this plateau appears only 
  for cut-offs beyond 3000~cm$^{-1}$. At lower energy, the metallic contribution is the dominant mechanism 
  for spectral weight redistribution. 
 If this observation was to be related to the onset of the pseudogap, it would imply 
 that the spectral weight redistribution 
extends far beyond our full experimental range, unlike PCCO (in fig.2, right panel,  
the spectral weight is  eventually conserved at 20000~cm$^{-1}$).
 A further concern is that in underdoped BSCCO single crystals, this plateau is not seen \cite{Carbone06}.
Hence its possible interpretation in terms of a pseudogap certainly requires one more step in the analysis. 

\section{Discussion}
%\label{3}

Having shown that the overdoped samples display a metallic behavior, we propose to normalize, 
 within the experimental error,  the 
spectral weight of the underdoped sample to the  overdoped one at 300~K.
This is shown in fig.5  for $\Omega=$8000~cm$^{-1}$, for the sake of a comparison with single crystals. 
Similar plots can be found for PCCO samples in earlier papers \cite{Zimmers05,NBAnnals}.
After normalizing so,  both samples appear to have a similar metallic 
behavior down to $\sim$~140~K,   within the error bars.
Next we superimpose single crystals data at $\Omega=$10000~cm$^{-1}$ \cite{Carbone06}, 
also normalizing $\rm N_{eff}$(300~K) of the underdoped crystal to the overdoped one. 
These two different sets of data turn out to be in surprisingly good agreement. Moreover,  
the spectral weight increases less rapidly in the underdoped sample than in the overdoped one, 
below approximately the same temperature. The difference develops  and reaches 
 $\sim$~0.005, still larger than our error bar. 
 Therefore, in this respect, we observe "missing spectral weight" below $\sim$~140~K. 
 We suggest that it is  a (subtle) signature of the onset of the pseudogap in underdoped BSCCO.

According to the terminology recalled above, the pseudogap in BSCCO is a "low energy" one, 
whereas in PCCO it would be a "high energy" one.  
If  such a distinction is valid, then different mechanisms for the onset of these pseudogaps may very well be 
at stake, as suggested by numerous authors. These pseudogaps might however be related: in hole doped cuprates, 
the broad feature in ARPES tracks the low energy pseudogap \cite{ZXrev}). Our  contribution  here is to show 
that both have an optical signature in terms of missing spectral weight, although the signatures are 
qualitatively and 
quantitatively different. More accurate studies of the in-plane conductivity in hole-doped materials are still 
necessary in order to allow a more comprehensive comparison with the pseudogap in n--type materials.  

%-------------
\begin{figure}
 %\begin{center}
      \includegraphics[scale = 0.45]{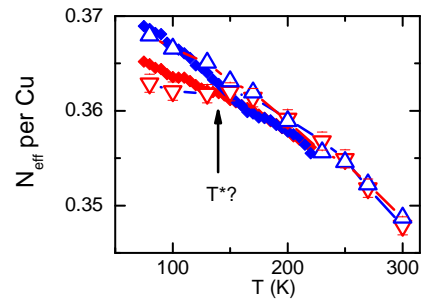}
 % \end{center}
  \caption{ Spectral weight computed for $\Omega$ = 8000~cm$^{-1}$ (films, open symbols \cite{Santander04}) 
  and 10000~cm$^{-1}$   (crystals, full symbols \cite{Carbone06}). 
  For each underdoped sample,  the absolute value has been normalized so as to overlap the 300~K 
  spectral weight of the  overdoped akin compound. The spectral weight of underdoped samples depart 
  form the overdoped 
  at $\sim$ 140~K.}    
  \label{Fig4}
\end{figure}
%-------------

\section{Conclusion}
%\label{4}

From the temperature dependence of the spectral weight in n-type PCCO cuprates, one can clearly 
see that the Fermi surface  breaks off into arcs  due to the opening of a pseudogap.
  In p-type BSCCO samples, the pseudogap  does not give rise to such a 
clearcut signature.  We used the metallic temperature  variation of the overdoped sample as a reference 
variation. Doing so, the temperature dependence for the  underdoped samples departs at a temperature 
$T^{\star}$ from the reference temperature dependence. This difference is large for PCCO (10~\%), 
but does not exceed 1.5~\% in terms of the total spectral weight in BSCCO.
We suggest that in the latter compound, it is a subtle  signature of the pseudogap. 
Whether the difference relates to a basic one {\it i.e.} 
a different mechanism for these partial gaps, remains to be settled.    

{\bf Acknowledgments}

The author is grateful to R.~Lobo, A.F.~Santander, A.~Zimmers whose work has been invaluable, and 
to L.~Benfatto, J.~Carbotte,  A.~Millis, T.~Timusk for illuminating dicussions. 
High quality samples have been grown and kindly provided by H.~Raffy's group in Orsay (BSCCO) 
and  R.~Greene's laboratory at the University of Maryland.


\begin{references}
\bibitem{Neff}  We  used a unit cell volume (per Cu) of 183~\AA$^3$ for PCCO, and 228~\AA$^3$ in 
BSCCO. The small differences in the  $\rm N_{eff}(300K)$  can depend on specific 
details of the technical procedure used to establish the absolute value of the spectral weight \cite{Santander04}.
\bibitem{note2} The accuracy is even better at lower energies $\sim$~10000~cm$^{-1}$, 
being further from our highest available measuring frequency.
\end{references}
\end{document}